\documentstyle[preprint,aps,floats,epsf]{revtex}

\tightenlines

\def\Eslash{\not{\hbox{\kern-4pt E}}}
\def\epem{e^+e^-}
\def\egm{e\gamma}

\def\et{\tilde e}

\def\ellt{\tilde \ell^-}
\def\chit{\tilde \chi^0_1}
\def\nut{\tilde \nu}

\begin{document}

\preprint{
\vbox{\hbox{\bf MADPH-98-1076}
      \hbox{\bf hep-ph/9808466}
      \hbox{August, 1998}}}

\title{ Slepton Oscillation at $\egm$ Colliders}

\author{J.-J. Cao$^{a,c}$, T. Han$^{b}$,  X. Zhang$^a$, and G.-R. Lu$^c$}

\address{$^a$Institute of High Energy Physics, Academia Sinica,
Beijing 100039, P. R. China\\
$^b$Department of Physics, University of Wisconsin, Madison, WI 53706, USA\\
$^c$Department of Physics, Henan Normal University, Xinxiang, Henan, P.R.
China}

\maketitle
\thispagestyle{empty}  

\begin{abstract}

We study the possible signature for lepton-flavor
violation processes in the scalar-lepton sector of
the minimal supersymmetric model at future electron-photon colliders. 
We find that an $ \egm $ collider can provide a good
opportunity to probe the mass differences and 
flavor mixing angles between the sleptons of the first
generation and the other two generations for the 
right-handed as well as the left-handed sector.
The sensitivity reach for lepton-flavor
violation is comparable to that obtained at an $e^+e^-$
and $e^-e^-$ collider, and is
significantly better than the current bound
from low-energy rare processes
such as $ \mu \rightarrow e \gamma $.
\end{abstract}

\newpage
\section{Introduction}

Supersymmetry (SUSY) is believed to be 
the most promising candidate for physics beyond the 
Standard Model (SM).
Among many attractive features, supersymmetry
can provide understanding for the mass hierarchy
of the weak scale and the Planck scale.
The existence of the SUSY particles (sparticles) 
at the weak scale can satisfactorily
lead to the gauge coupling unification at a desirable scale.
The electroweak symmetry breaking 
is triggered radiatively by the large top-quark mass.
The weak-scale SUSY also necessarily leads to rich
physics in the near-future collider experiments \cite{susy}.

However, current supersymmetric models do not advance 
our knowledge of flavor physics beyond the SM, such as 
the fermion mass generation and quark/lepton mixings. 
In fact, 
generic SUSY models often have arbitrary flavor mixings 
and mass parameters in the scalar quark (squark) and scalar
lepton (slepton) sectors 
and one would have to assume certain symmetries to 
prevent large flavor-changing neutral currents 
(FCNC) \cite{fcnc}.  
On the other hand, the flavor structure
in the SUSY sector motivates us to seek for new physics, 
and any experimental observation on the FCNC 
processes beyond the SM would undoubtedly shed light on 
our understanding for flavor physics.
The evidence of neutrino oscillations observed by the
Super-Kamiokande experiments serves as a good 
example \cite{SuperK}.

Besides the conventional studies for FCNC effects in SUSY
grand unification models in processes 
$b\to s\gamma$ \cite{btosgm}, 
neutral meson oscillations \cite{meson},
and $\mu \to e\gamma$ \cite{mu-eg} at low energies,
it has been actively pursued to explore the possible
signatures at future collider experiments for the 
lepton-flavor violating processes in SUSY theories. Due to
the super-GIM mechanism, it has been pointed out \cite{flavor,CP} 
that for low-energy rare 
processes induced by sparticle loops, the rates are suppressed 
by $\Delta M/M$; while for the flavor changing signal from direct
slepton production at high energy colliders the suppression
factor is at most $\Delta M/\Gamma$, where $M$ and $\Gamma$ are
the sfermion mass and decay width.
It is found that there is great physics 
potential for the CERN Large Hadron Collider
(LHC) and the Next $\epem$ Linear Collider
(NLC) to probe SUSY flavor physics through slepton
oscillations \cite{flavor,CP}.
In contrast, there are several advantages for considering 
an $\egm$ collider to study SUSY flavor physics.
First of all, lepton and electron-photon colliders
often provide much cleaner experimental environment
for new physics searches than hadron colliders. 
Second, in models with soft mass unification or 
with (minimal) gauge-mediation of SUSY breaking, 
the lightest neutralino ($\tilde\chi_1^0$) is expected 
to be lighter than sleptons ($\tilde \ell$). 
Therefore, the sparticle production for
$e^\pm \gamma\to \tilde \ell\tilde\chi^0,\  \nut \tilde\chi^\pm $
would have lower mass threshold \cite{egm0,kongoto,bhk} 
than those for $\epem \to \ellt \tilde\ell^+, \nut \nut^*$.
Third, even when slepton pair production is kinematically allowed
there is a $P$-wave ($\beta^3$) suppression of the 
cross section near threshold,
and the event rates are correspondingly limited.
On the other hand, the cross sections for  
$\tilde \ell\tilde\chi^0,\  \nut \tilde\chi^\pm$
production are proportional to
$\beta$ near threshold, so high production rates are achievable.
Fourth, since there exists the net electron flavor in the initial 
state, the observation on a lepton flavor other than an electron
in the final state could be a clear signature for the flavor 
oscillation, with less severe SM backgrounds.

In this paper, we study the signals for slepton oscillation
between the first generation and the other two generations
at an $e\gamma$ collider and their SM backgrounds.
We consider an $e^+e^-$ collider with 500 GeV c.~m.~energy 
and an annual luminosity 50 fb$^{-1}$ \cite{NLC},
running at the $e\gamma$ mode \cite{telnov}. In section II, 
we first illustrate the general flavor mixing in the
slepton sector for the minimal supersymmetric standard
model (MSSM). We then present the slepton production 
mechanisms and cross sections with and without the
flavor mixing. In section III, we study the slepton
oscillation signals and their SM backgrounds at an 
$e\gamma$ collider. 
We find that for the gaugino-like neutralino and chargino,
it is promising to probe the mass differences and mixing angles 
between the left-handed sleptons to good precision. 
For the right-handed slepton sector, the results are
comparable to the existing one at 
$e^+e^-$ and $e^-e^-$ linear colliders \cite{flavor}.
The sensitivity obtained 
here is better than the most stringent low-energy constraints
for lepton-flavor violation from $\mu \to e \gamma$ \cite{mu-eg}.
We summarize our results in Sec. IV.

\section{Lepton-Flavor Oscillations at $\egm$ Colliders}

The mass terms for the left-handed sleptons (partners to the
left-handed leptons), right-handed sleptons and left-handed 
sneutrinos can be generally written as
\begin{equation}
\et^*_{L\alpha} M^2_{L\alpha \beta} \et_{L\beta} +
\et^*_{R\alpha} M^2_{R\alpha \beta} \et_{R\beta} + 
\nut_{\alpha}^* M^2_{\nut \alpha \beta} \nut_{\beta},
\end{equation}
where $\alpha, \beta=1,2,3$ are the generation indices, 
and $M_{L,R,\nut}$ are mass matrices
which can be diagonalized through the unitary matrices 
$W_{L,R,\nut}$. In terms of the slepton mass eigenstates, 
the lepton-slepton interactions
with gauginos are flavor non-diagonal: 
\begin{equation}
\et_{Li} W^*_{Lij}  {\overline e_{Lj}} \tilde{\chi}^0 +
\et_{Ri} W^*_{Rij}  {\overline e_{Rj}} \tilde{\chi}^0 +
\nut_i W^*_{\nut ij}{\overline e_{Lj}} \tilde{\chi}^- +h.c. ,
\label{fcnc}
\end{equation}
which generally leads to lepton-flavor violating effects.
If the slepton mass differences are large and
the lepton-flavor off-diagonal couplings ($W_{ij}$) 
are sizeable, Eq.~(\ref{fcnc}) may directly lead to 
flavor-changing processes like
\begin{eqnarray}
\label{Process}
e^- \gamma & \to & \nut_\ell \tilde{\chi}^-_i, \\
e^- \gamma & \to & \ellt \chi^0_j, 
\label{Process1}
\end{eqnarray}
where the flavor index $\ell=\mu, \tau$, as depicted in Fig.~1(a) and (b), 
in which the black dots denote the vertices with flavor changing
interactions. On the other hand, 
if $\et^-$ ($\nut_e$) is nearly degenerate with other sleptons
(sneutrinos), as prefered by suppressing the FCNC and as
constrained by $\mu \to e \gamma$ data,  
then the processes (\ref{Process}) 
and (\ref{Process1}) may mainly go
through the flavor diagonal production  
$ \et^- \chi^0_j (\nut_e \tilde{\chi}^-_i) $ followed
by $ \et^- \to \ellt ( \nut_e \to \nut_\ell\ )  $ 
oscillation subsequently \cite{flavor}.
The two cross sections are then related by
\begin{eqnarray}
\label{osci}
\sigma(e \gamma \to \ellt \chit)& = &\sigma (e \gamma \to \et^- \chit) 
P(\et^- \to \ellt), \\ 
P(\et^- \to \ellt) & =  & 
2 \sin^2{\theta_\ell} \cos^2{\theta_\ell}
\frac{(\triangle M^2)^2}{4 {\overline m}^2 \Gamma^2+ (\triangle M^2)^2},
\nonumber
\end{eqnarray}
where $P(\et^- \to \ellt)$ is the transition probability 
of interaction eigenstates $\et^-$ to $\ellt$ via oscillation,  
and $\triangle M^2 =m_{\tilde l}^2 -m_{\et}^2,\  
\overline{m}=(m_{\tilde l}+m_{\et})/2,$ and $\theta_\ell$ 
the mixing angle between $\et^-$ and $\ellt$.
The cross section formula for sneutrino oscillation 
is the same as Eq.~(\ref{osci}) with the replacement of 
$\tilde{e} \rightarrow \nut_e, \ellt \rightarrow \nut_{l}$.

For the flavor diagonal production of $\nut_e \tilde{\chi}^-$, 
the differential cross section 
summed over the chargino helicity are \cite{bhk}
\begin{eqnarray}
&&\hspace{0.1in}\frac{d\sigma}{d\cos\theta}(e^-_- \gamma_- \to 
\tilde{\nu}_{eL}
\tilde\chi^-_j) =
{\pi\alpha^2 \over \sin^2{\theta_W}} \ {V_{j1}^2\ \over s} \
\frac{r^2_{\tilde \chi}}{(1-\beta^2)} \frac{\beta}{(1+\beta\cos\theta)^2}
\nonumber\\
&&\hspace{0.3in}
\times \ \sum_{\lambda =\pm 1} (1+\lambda\cos\theta) (1+\lambda\beta)^2
\left(\frac{\sqrt{1-\beta^2} }{r_{\tilde \chi}} - (1+\lambda\beta)\right),
\label{nu1}\\
&&\hspace{0.1in}\frac{d\sigma}{d\cos\theta}(e^-_- \gamma_+ \to\tilde
{\nu}_{eL}
\tilde\chi^-_j) = 
{\pi\alpha^2 \over \sin^2{\theta_W}}\ {V_{j1}^2 \over s}\
\frac{r^2_{\tilde \chi}}{(1-\beta^2)}
\ \frac{2\beta^3\sin^2\theta (1-\beta\cos\theta)^2 }
{(1+\beta \cos{\theta})^2}. 
\label{nu2}
\end{eqnarray}
The subscripts on $ e $ and $ \gamma $ refer to the electron and photon
helicities. The angle $ \theta $ specifies the chargino momentum relative to 
the incoming direction in the c.m. frame, $ \beta=p/E $ is the chargino
velocity in the c.m. frame, and 
$ r_{\tilde \chi} =m_{\tilde \chi} /\sqrt{s}$. 
The $(\lambda_e , \lambda_{\gamma})=(-,-)$ helicity amplitude is
$S$-wave
near threshold so the cross section of Eq. (\ref{nu1}) is proportional to 
$ \beta $; the $(-, +)$ helicity amplitude, which comes only from the
$t$-channel diagram, is $P$-wave near threshold so the cross section of
Eq. (\ref{nu2}) goes like $\beta^3$. 
the scattering amplitude is proportional to the wino
fractions $ V_{j1} $ of the matrix $ V_{ji} $ that diagonalizes the
mass matrix  (the first index $j$ labels the chargino mass  eigenstate 
$ \tilde{\chi}^+_1, \tilde{\chi}^+_2 $ and the second index $ i $ refers 
to the primordial gaugino and Higgsino basis $ \tilde{W}^{\pm}$, $
\tilde{H}^{\pm} $). Further, the $ \nut_L $ state fixes the incoming
electron chirality to be left-handed ``$-$''.

Selectron-neutralino associated production 
$e\gamma\to\tilde e\tilde\chi^0$ 
proceeds via $ s $-channel electron and $t$-channel
selectron exchanges \cite{egm0,bhk}; see Fig. 1(b) with 
$\tilde{\ell}=\tilde{e}$. By Eq.~(\ref{osci}), 
we can get the cross section of $e \gamma \to \tilde{\ell}
\tilde{\chi}^0 $. The contributions from Higgsino components
($\tilde H_1^0, \, \tilde H_2^0$) of $\tilde\chi^0$
can be neglected and only the neutralino mixing
elements $Z_{j1}$ and $Z_{j2}$ enter.
After summing over the neutralino helicities,
there are four independent helicity cross sections as the
helicity of the
$\tilde e_R$ ($\tilde e_L$) matches that of the $e_R$ ($e_L$):
\begin{eqnarray}
&&\hspace{0.1in}\frac{d\sigma}{d\cos\theta}(e^-_+ \gamma_+\to\tilde
e_{R}^-
\tilde\chi^0_i) =\frac{d\sigma}{d\cos\theta}(e^-_- \gamma_-\to\tilde
e_{L}^-\tilde\chi^0_i)=
\pi\alpha^2 \ {2F_{i(L,R)}^2\ \over s} \
\frac{r^2_{\tilde e}}{(1-\beta^2)}
\nonumber\\
&&\hspace{0.3in}
\times \frac{\beta}{(1+\beta\cos\theta)^2} \
\sum_{\lambda =\pm 1} (1+\lambda\cos\theta) (1+\lambda\beta)^2
\left(\frac{\sqrt{1-\beta^2} }{r_{\tilde e}} - (1+\lambda\beta)\right),
\label{neuthel1}\\
&&\hspace{0.1in}\frac{d\sigma}{d\cos\theta}(e^-_+ \gamma_-\to\tilde
e_{R}^-
\tilde\chi^0_i)  = \frac{d\sigma}{d\cos\theta}(e^-_- \gamma_+\to\tilde
e_{L}^-\tilde\chi^0_i)=\nonumber\\
&&\hspace{0.4in}
\pi\alpha^2\ {2F_{i(L,R)}^2 \over s}\ \frac{r^2_{\tilde e}}{(1-\beta^2)}
\ \frac{2\beta^3\sin^2\theta}{ (1+\beta\cos\theta)^2 }
\left(\frac{\sqrt{1-\beta^2}}{r_{\tilde e}} -
(1-\beta\cos\theta)
\right).\label{neuthel2}
\end{eqnarray}
The $F_{iL}$ $(F_{iR})$ for $\tilde e_L$ $(\tilde e_R)$
are effective couplings given by
\begin{eqnarray}
F_{iL} = \frac{1}{2}\left[\frac{Z_{i1}}{\cos\theta_W}
+\frac{Z_{i2}}{\sin\theta_W}\right], \quad
F_{iR} = \frac{-Z_{i1}^{*}}{\cos\theta_W}.
\label{lrcouplings}
\end{eqnarray}
Here the $Z_{ji}$ are the elements of matrices that diagonalize
the neutralino mass matrix (the first index $j$ labels the neutralino
mass eigenstate $\tilde\chi_j^0$, $j=1...4$, and the second index
$i=1,2$ refers to the primordial gaugino and Higgsino basis
($\tilde B^0,\, \tilde W^0,\, \tilde H_1^0,\,\tilde H_2^0$)).
The angle $\theta$ specifies the direction of the selectron
with respect to the direction of the incoming electron in the c.m.\
frame,
$\beta$ is the velocity of the selectron, and
$r_{\tilde e}=m_{\tilde e} /\sqrt{s}$. 

Due to the approximate decoupling of Higgsinos from the electron,
the cross sections for the processes (\ref{Process}) and
(\ref{Process1})
are only large when the charginos and neutralinos
are mainly gaugino-like, namely
$\tilde\chi^\pm \sim \tilde W^\pm$ and
$\tilde\chi^0_1\sim \tilde B^0,\ \tilde\chi^0_2 \sim \tilde W^0$.
Fortunately gaugino-like
$\tilde\chi_1^\pm$, $\tilde\chi^0_1$ and $\tilde\chi^0_2$ are
highly favored theoretically
for two reasons: ({\it i}) the radiative electroweak symmetry
breaking in SUSY GUTs theories yields a large $|\mu|$ value if
$\tan\beta$ is bounded by the infrared fixed point solutions
for the top quark Yukawa coupling \cite{tanbeta,susygut};
({\it ii}) $\chi_1^0\sim\tilde B$ is strongly preferred 
for $\chi_1^0$ to be	
a viable cold dark matter candidate \cite{tanbeta,susygut,dm}.
In the rest of our paper,
we will thus concentrate on this scenario.
For our illustrations we choose the
chargino/neutralino masses (in GeV)
\begin{equation}
m_{\tilde\chi_1^0} = 64, \qquad m_{\tilde\chi_2^0} = 130, \qquad
m_{\tilde\chi_1^\pm} = 130, \qquad m_{\tilde\chi_2^\pm} = 294,
\end{equation}
and the mixing matrix elements
\begin{equation}
Z_{11} = 0.95, \qquad Z_{12} = -0.20, \qquad Z_{21} = -0.28, \qquad
Z_{22} = -0.90, 
\end{equation}
for the neutralinos and 
\begin{equation}
V_{11} =0.96, \qquad V_{21}=-0.27,
\end{equation}
for the chargino.  
These parameters correspond to the following MSSM
parameters at the weak scale
\begin{equation}
M_1 = 62 {\rm\ GeV},\quad M_2 = 126\ {\rm GeV}, \quad
\mu = 265 \rm\ GeV\ , \quad \tan\beta=3\ ,
\label{masstanb}
\end{equation}
where $\tan\beta$ is a little larger than that at the infrared fixed point
value\cite{tanbeta}  to get a heavier Higgs mass to avoid the LEP2 limit
and the
convention
for sign$(\mu)$ follows Ref.~\cite{tanbeta}.
For slepton masses, we choose (in GeV)
\begin{eqnarray}
m_{\tilde \mu_L} \approx m_{\tilde e_L} = 275,  \qquad  
m_{\tilde \mu_R} \approx m_{\tilde e_R} = 260,  \qquad  
m_{\tilde\nu_\mu}\approx m_{\tilde \nu_e} = 266.
\end{eqnarray}
Our choices for the gaugino and slepton masses
are consistent with
renormalization group evolution \cite{susygut}
to the electroweak scale,
with the following universal mSUGRA parameters
\begin{eqnarray}
 m_0 = 250 {\rm\ GeV}\,,\qquad
 m_{1/2} = 150 {\rm\ GeV}\,,\qquad
 A = 0 \,.
\label{gutmass}
\end{eqnarray}

From Eqs. (\ref{nu1}--\ref{neuthel2}), 
we see the following features:
Due to the stronger diagonal couplings $Z_{ii}$,
the cross sections for $\tilde e_L \tilde\chi_2^0$
(mainly $\tilde e_L \tilde W^0$) and
$\tilde e_R \tilde\chi_1^0$ (mainly $\tilde e_R \tilde B^0$)
are significantly larger than the other selectron channels;
the weaker neutral current couplings and the more massive
scalar propagator in selectron production make their cross
sections smaller than those for sneutrino production. 
We illustrate these features in four processes with
larger cross sections in Fig.~2. Two non-zero $e,\gamma$ 
helicity combinations are labeled in parentheses. 
The solid curves are of $S$-wave near the threshold
resulting in a higher cross section rate,
and the dashed ones of $P$-wave. This
indicates the possible signal enrichment by properly
choosing the helicity configuration for the beams. 
The lower curves show the effects of convolution
with the backscattered laser photon spectrum for the machine
parameters as chosen in Refs.~\cite{kongoto,bhk}.  
The effect is to decrease the
cross sections by about a factor of two. At an NLC with 
c.~m.~energy 500 GeV, the typical production sections are of order
1000 fb for $e\gamma \to \tilde \nu_e \tilde\chi_1^-$, and
about 100 fb for $e\gamma \to \tilde e_L \tilde\chi_2^0,$ and 
$\tilde e_R \tilde\chi_1^0$, on which three channels
we will focus for our phenomenology study in the next section.

\begin{table}[t]
\bigskip
\caption{\label{br-bf}  
Sparticle decay modes and branching fractions for the
representative parameter choice. Here $q$ generically denotes a quark
and $l$ denotes a lepton; fermion-antifermion pairs $ f\bar{f}^{\prime} $
with net charge $-1 (0)$ are denoted by $C^- (N^0)$}
\begin{center}
\begin{tabular}{cc}
Decay Modes  & Branching Fraction (\%)\\
\hline
$\tilde{\chi}_1^- \to {\tilde{\chi}_1^0} C^-$ &  79.5 ($\tilde{\chi}^0_1 q
\bar{q}^{\prime} $),\  20.5 ($ {\tilde{\chi}^0_1} l^- \nu $)  \\
\hline
 $\tilde{\chi}_2^0 \to \tilde{\chi}_1^0 N^0 $ & 72 ($ \tilde{\chi}^0_1 q
\bar{q}$ ), 19 ($\tilde{\chi}^0_1 \nu \bar{\nu} $), 9 ($\tilde{\chi}^0_1
{l^+ l^-}$)\\
\hline
$ \tilde{\mu}_R \to \tilde{\chi}_1^0 \mu^- $ & 100\\   
\hline
$\tilde{\mu}_L \to \tilde{\chi}^-_1 \nu,\ \tilde{\chi}^0_2 \mu^-,\
\tilde{\chi}^0_1 \mu^- $ & 54, 28, 18 \\
\hline
$ \tilde{\nu}_{\mu} \to \tilde{\chi}^+_1 \mu^-,\ \tilde{\chi}^0_2 \nu, \
\tilde{\chi}^0_1 \nu $  & 61, 29, 10 \
\end{tabular}
\end{center}
\end{table}

\section{Signal Selection and Background Suppression}

We first consider the $e-\mu$ flavor oscillation.
The muon slepton and sneutrino production rates can be obtained by
Eqs.~(\ref{osci}--\ref{neuthel2}). Before considering the experimental 
signature at an $e\gamma$ collider, in Table~\ref{br-bf} we illustrate the 
predicted branching fractions of the final state sparticle decays
\cite{isasugra} for the parameters discussed earlier.
We have introduced notations $C^\pm,N^0$ to denote the
net charge of the final $f\bar f'$. They are largely from
$W^{*\pm},Z^{*0}$ decays, so that we can more directly
compare with SM background processes.
The decay widths for the sleptons  are also an important 
variable in the calculation of Eq.~(\ref{osci}). 
They are calculated to be 
$\Gamma_{\tilde{\ell}_R}=1.2$ GeV, $\Gamma_{\tilde{\ell}_L}=2.6$
GeV and $\Gamma_{\tilde{\nu}_L}=2.1$ GeV for the parameters 
discussed in the last section.
The neutrinos and ${\chi}^0_1$ result in missing energies
in the detector; while the light quarks will lead to hadronic jets.
Combining the sparticle production and decay, 
we classify the final state signatures in Table~\ref{process-bf}
with corresponding branching fractions. 
The decays of the sparticles in these reactions 
generally give distinctive signals: 
Large missing energy ($\Eslash$), energetic charged
leptons or jets from light quarks. 
To effectively suppress the SM background, 
we utilize the following channels:
\begin{equation}
e^- \gamma \to \tilde{\nu}_{\mu} \tilde\chi_1^- \to \mu^-  \Eslash \ 
q \bar{q}^{\prime} \ q \bar{q}^{\prime},  \label{channel1}  
\end{equation}
to probe the mixing angle $ \theta_{\nu L} $ between $ \tilde{\nu}_e $ and
$ \tilde{\nu}_{\mu} $; the process
\begin{eqnarray} 
e^- \gamma \to \tilde{\mu}_L \tilde{\chi}_2^0 \to 
\mu^- \Eslash \ q \bar{q} \ q \bar{q},
\label{channel2}
\end{eqnarray}
for $\theta_L $ between $ \tilde{e}_L $ and $ \tilde{\mu}_L $ and the
process
\begin{eqnarray}
e^- \gamma \to \tilde{\mu}_R \tilde\chi_1^0 \to \mu^- \Eslash,
\label{channel3}
\end{eqnarray}
for $ \theta_R $ between $ \tilde{e}_R $ and $ \tilde{\mu}_R $.

\begin{table}[t]
\bigskip
\caption{\label{process-bf} Sparticle production and decay
in $e^-\gamma$ collisions. Branching fractions based on
Table~\ref{br-bf} are given in the parentheses.
Here $\Eslash$ denotes missing energy resulting from
$\tilde\chi_1^0$ and $\nu$ final state; $C^-$ ($N^0$)
denotes a fermion-antifermion pair of charge $-1$ ($0$).}
\begin{center}
\begin{tabular}{cc}
Process & Final State \& Branching Fraction\\
\hline 
$e^- \gamma \to \tilde\nu_{\mu} \chi^-_1 \to $ & $ C^- \Eslash\ (16\%),\quad
C^-N^0 \Eslash\ (23\%),\quad C^-C^+ \mu^- \Eslash\  (61\%) $ \\
\hline
$e^- \gamma \to \tilde \mu_L \tilde\chi_2^0 \to$
& $ N^0 \mu^-\ \Eslash$\ (23\%),\quad $ N^0 \ N^0 \mu^-\ \Eslash$\  (18\%),
\quad $C^- \ N^0 \ \Eslash$\ (44\%),\quad $C^- \ \Eslash$\ (15\%) \\
\hline
$e^- \gamma \to \tilde \mu_R \tilde\chi_1^0 \to$
& \ $ \mu^- \Eslash$ \ (100\%) 
\end{tabular}
\end{center}
\end{table}

Generally speaking, due to the distinctive kinematical characteristics,
the multiple-jet ($q \bar q'$) signals for $\tilde\nu_{\mu} \tilde\chi_1^-$
and $ \tilde{\mu}_L \chi_2^0 $ have favorable signal-to-background ratios.
The irreducible SM background processes are
\begin{equation}
e^- \gamma \to  \mu^- \nu_\mu\ W^-W^+,\quad  \mu^- \nu_\mu\ ZZ.
\label{back}
\end{equation}
Including the hadronic decay branching fractions of $W,Z$ decays,
we find that the cross sections of Eq.~(\ref{back}) are
less than 0.1 fb at $\sqrt s_{ee}=$ 500 GeV.  We could therefore
expect a very effective probe to the oscillation parameters
for processes (\ref{channel1}) and (\ref{channel2}).
However, the leading background to the 
$\tilde \mu_R^{} \tilde\chi_1^0$ signal of Eq.~(\ref{channel3}) is
\begin{equation}
e^- \gamma \to \mu^- \bar \nu_\mu \nu_e.
\label{big}
\end{equation}
The total cross section for this process at $\sqrt s_{ee}=$ 500 GeV
is about two picobarns, and the oscillation signal would
be swamped by it. On the other hand, this signal process involves
a right-handed electron in the initial state while the background 
process involves only a left-handed one due to the
$W$ exchange. If high beam polarization $e^-_R$ can be 
implemented \cite{NLC}, 
then this SM background can be highly suppressed. 
Helicity argument shows that choosing the right-handed photon 
beam could also help further suppress the background survived
from the impure $e^-_L \gamma_+$ reaction, without hurting the
signal [see Fig. 2(b)]. We thus assume that the background 
Eq.~(\ref{big}) can been reduced to a manageable level.
In our subsequent numerical discussions,
based on the signal branching ratios in Table~II,
we estimate the signal efficiencies to be 40\%, 15\% and 80\% for 
Eqs.~(\ref{channel1}), (\ref{channel2}) and (\ref{channel3}), 
respectively. We consider an oscillation signal to be observable 
if the cross section after the inclusion of the efficiencies to be 
no less than 0.1 fb. For instance, a 0.1 fb signal cross section
and negligible backgrounds with a 50 fb$^{-1}$ luminosity would
correspond to about 99\% C.L. discovery.

In Figs.~3, 4 and 5, we illustrate contours of constant cross
sections for signal processes in (\ref{channel1}), (\ref{channel2}) 
and (\ref{channel3}), respectively, in 
$\triangle M^2-\sin{2 \theta_\ell}$ 
plane. When plotting each figure, 
we assume that just one mixing angle exits, {\it i.e.}, 
the contributions originating from other mixing angles
are set to zero. 
The (thick) solid curves in the figures present the constant
cross section contours for $\sigma=0.1,\ 0.5$ and 1 fb.
The dashed lines correspond to the current bound on the 
branching fraction $B(\mu \to e \gamma)<4.9\times 10^{-11}$.
A lower value of $4.9\times 10^{-12}$ is also indicated by
the (thin) solid line. Those curves are obtained with the
SUSY parameters presented in the previous section.
As showed in Figs. 3 and 4, achievable sensitivity in
$\triangle M^2 - \sin{2 \theta_\ell}$ for the left-handed slepton
sector can be significantly better than the low-energy
constraint from $\mu \to e \gamma$. It is even more impressive
to probe the oscillation parameters in the right-handed slepton 
sector if the beam polarizations can be implemented to
suppress the backgrounds. 
Compared with the resultes at $e^+e^-$ and $e^-e^-$
colliders \cite{flavor}, 
the potential of probing mixing angle $ \theta_R $ 
between $\tilde{\mu}_R$ and $\tilde{e}_R$ are comparable. 

So far, we have only discussed the slepton oscillations
between electron and muon flavors. 
Our analyses are essentially applicable for 
$\tilde{e}-\tilde{\tau}$ oscillation as well. In fact,
our calculations for the signal by Eqs.~(\ref{osci})--(\ref{neuthel2})
and the backgrounds of Eqs. (\ref{back})-(\ref{big}) 
should be formally identical 
for the $\tau$ final state. The only differences are: 
Theoretically, the mass difference between $\tilde{e}$ and 
$\tilde{\tau}$ may be bigger than that of  $\tilde{e}$ and  
$\tilde{\mu}$ due to the larger Yukawa coupling running 
for $\tilde{\tau}$. Experimentally, the $\tau$ lepton 
tagging is less efficient than that for $\mu$. This may
lead to about a 80\% reduction on the event rate. Nevertheless,
we expect about same order of magnitude sensitivity for
$\tilde{e}-\tilde{\tau}$ oscillation. Due to the electron 
flavor in the initial state, the signal processes under discussion
are not sensitive to the $\tilde{\mu}-\tilde{\tau}$ 
oscillation. Recent study \cite{mutau} showed that an 
$e^+e^-$ collider could probe this oscillation to high
precision.

\section{Discussions and Summary}

We only considered a representative set of SUSY parameters. 
As long as the lighter neutralino and chargino are gaugino-like
as anticipated in mSUGRA models and in the minimal gauge-mediated
SUSY breaking model, our results should be generally valid. We
presented our study with the slepton masses approximately 
270 GeV. This has been beyond the reach for an NLC with
$\sqrt{s_{ee}}=500$ GeV
via $\epem \to \ellt \tilde\ell^+, \nut \nut^*$.
For a heavier slepton spectrum, we would need a collider 
with a higher c.~m.~energy. The current discussions on
the linear collider parameters \cite{NLC} indicate the
feasibility to extend the c.~m.~energy to 1.5 TeV.
It is always most beneficial that the c. m. energy is just 
above the signal production threshold. However, if the 
neutralino and chargino are Higgsino-like, then their couplings
to leptons are too weak to result in significant
cross sections for the signal at an $e\gamma$ collider.

In summary, we have demonstrated that for gaugino-like 
neutralino and chargino, the
slepton oscillation effects may be observable at a 500
GeV $e\gamma$ collider with a 50 fb$^{-1}$ luminosity.
The achievable sensitivity in $\triangle M^2 - \sin{2 \theta_\ell}$ 
for the left-handed slepton sector
can be significantly better than the low-energy
constraint from $\mu \to e \gamma$. 
In the right-handed slepton sector, our results are comparable
to that for  $e^+e^-$ and $e^-e^-$ colliders,
if the beam polarizations ($e_R^-$ and $\gamma_+$)
can be implemented to suppress the backgrounds. 
It may even be possible for an $e\gamma$ collider to 
go beyond the probe at an $e^+e^-$ collider because of the
kinematics advantage. Our analyses are essentially applicable for 
$\tilde{e}-\tilde{\tau}$ oscillation, with a sensitivity
of the same order of magnitude as that for
$\tilde{e}-\tilde{\mu}$.

\section*{ Acknowledgments}

We thank H.-C.~Cheng and J.~Feng for discussions.
This work was supported in part by the U. S. Department of Energy
under Grant No.~DE-FG02-95ER40896 and by National Natural Science
Foundation of China. Further support was provided
by the University of Wisconsin Research Committee, 
with funds granted by the Wisconsin Alumni Research
Foundation.


\newpage

\begin{figure}
\centering\leavevmode
\epsfxsize=5in\epsffile{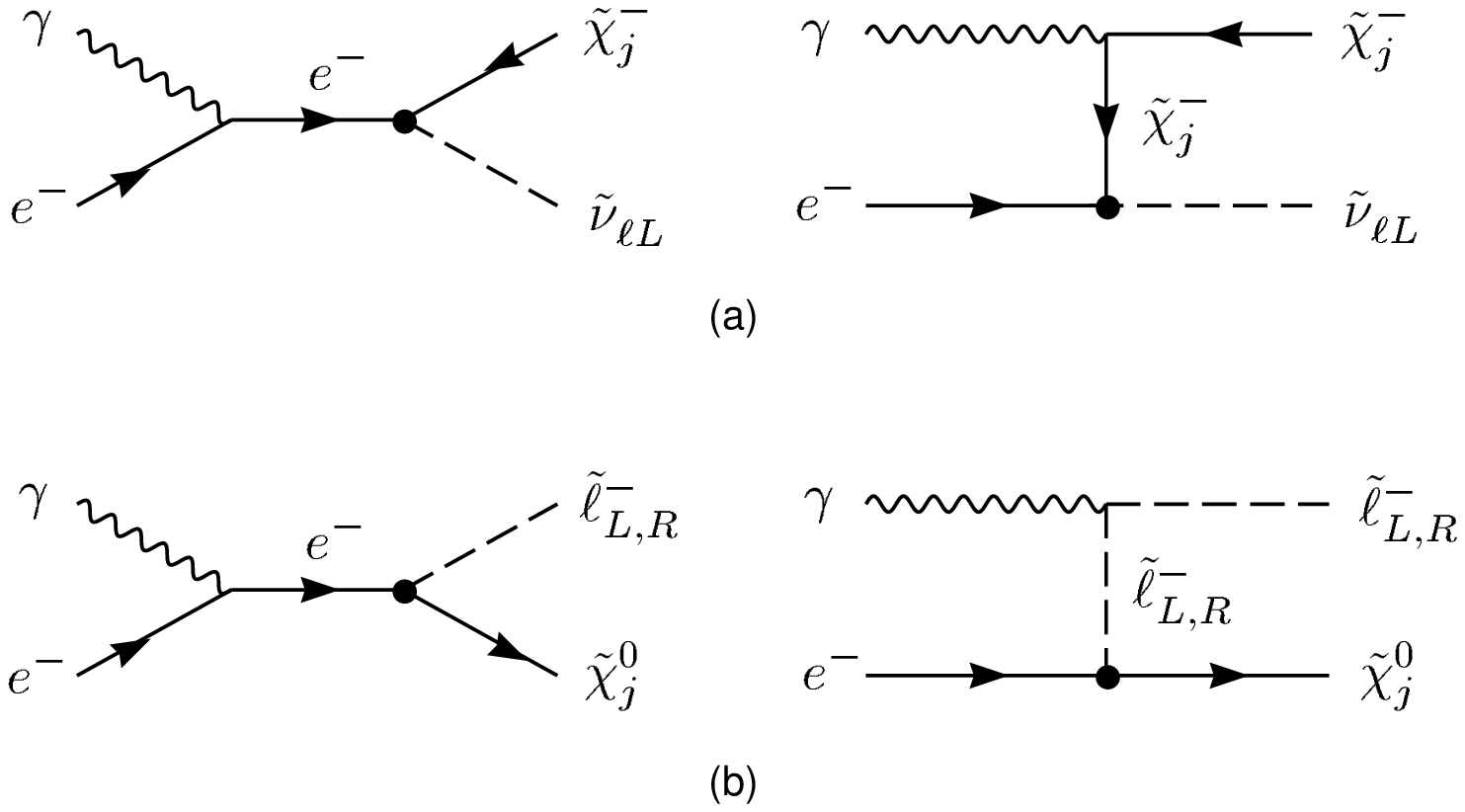}

\caption{Feynman graphs for slepton-gaugino associated production
(a) $ e^- \gamma \to \tilde{\nu_\ell} \tilde{\chi}^- $ 
and (b) $ e^- \gamma \to \tilde{l}^- \tilde{\chi}^0$. 
The black dots denote the new flavor violating vertices.}
\end{figure}

\begin{figure}
\centering\leavevmode
\epsfxsize=5in\epsffile{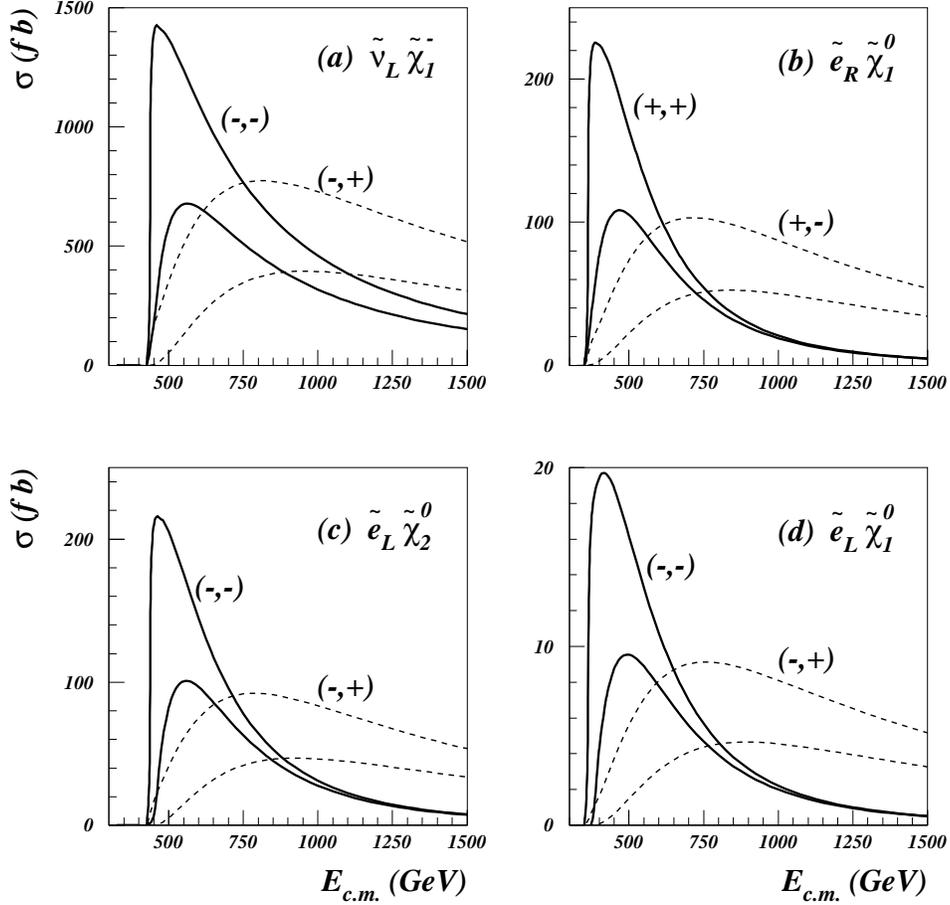}

\caption[]{Total cross section in fb versus the c.~m.~energy in GeV
for 
(a) $ e \gamma \to \tilde{\nu}_e \tilde{\chi}^-_1 $, 
(b) $ e \gamma \to \tilde{e}_R \tilde{\chi}^0_1 $, 
(c) $ e \gamma \to  \tilde{e}_L \tilde{\chi}^0_1 $ and  
(d) $ e \gamma \to  \tilde{e}_L \tilde{\chi}^0_2 $,
with the SUSY parameters given in the text. Electron and
photon beam polarizations are indicated in the parentheses.
The upper two curves are these at $e\gamma$ collider versus
$\sqrt{s_{e\gamma}}$.
The lower two curves are the corresponding results, convoluted with the
backscattered photon spectrum, versus $ \sqrt{s_{ee}} $. }
\end{figure}

\begin{figure}
\centering\leavevmode
\epsfxsize=5in\epsffile{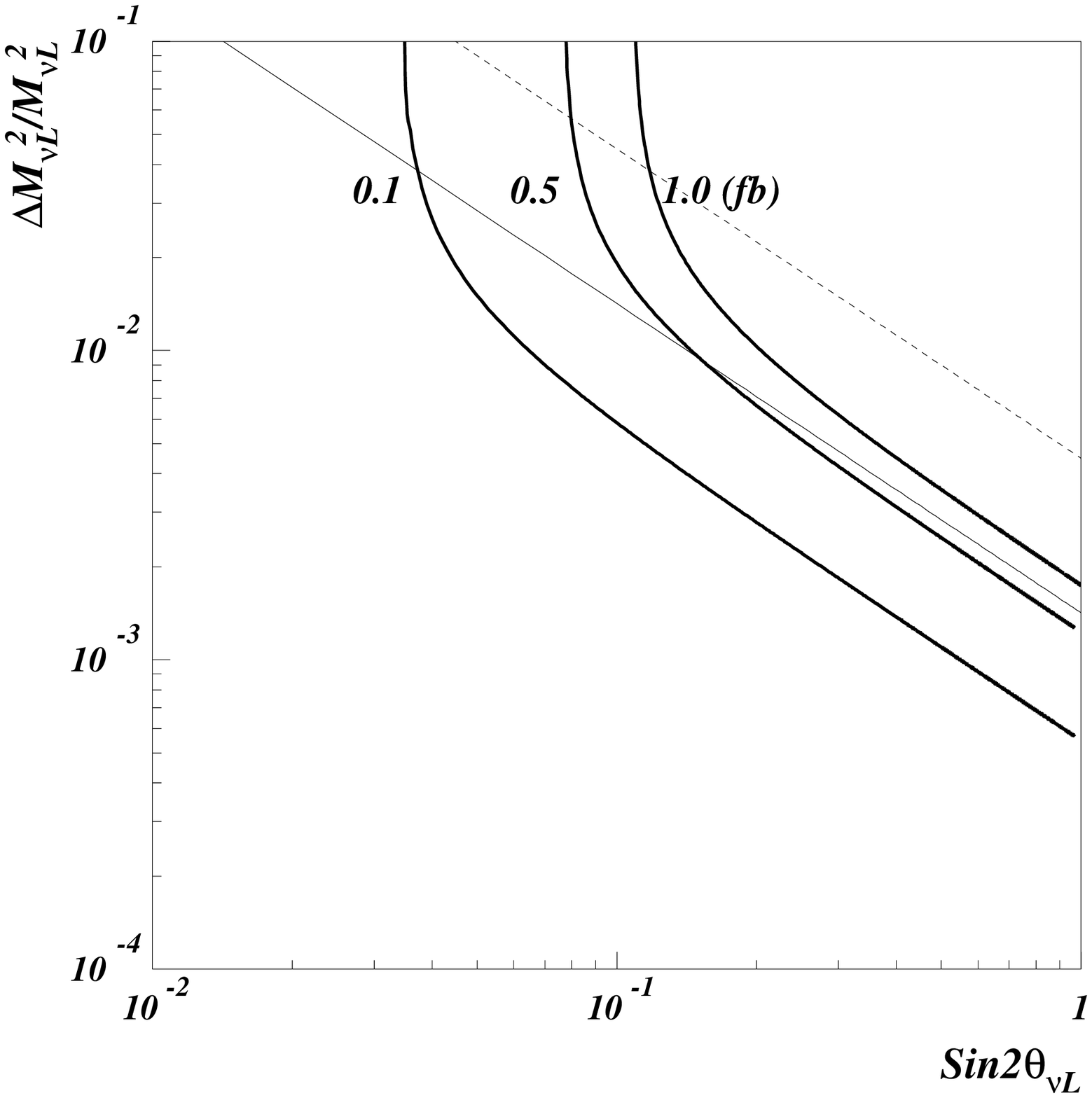}

\caption[]{Contours of constant cross sections for
$e^- \gamma \to  \tilde{\nu}_{\mu} \tilde\chi_1^- \to \mu^- \Eslash \ 
q \bar{q}^{\prime}\ q \bar{q}^{\prime}$ (thick solid)
with $ \sqrt{s_{ee}}=500 $ GeV and the SUSY parameters given
in the text. Constant contours of $ B(\mu \to e \gamma ) =4.9 \times 
10^{-11} $ (dotted) and $ 4.9 \times 10^{-12}$ (solid) are also plotted.}
\end{figure}

\begin{figure}
\centering\leavevmode
\epsfxsize=5in\epsffile{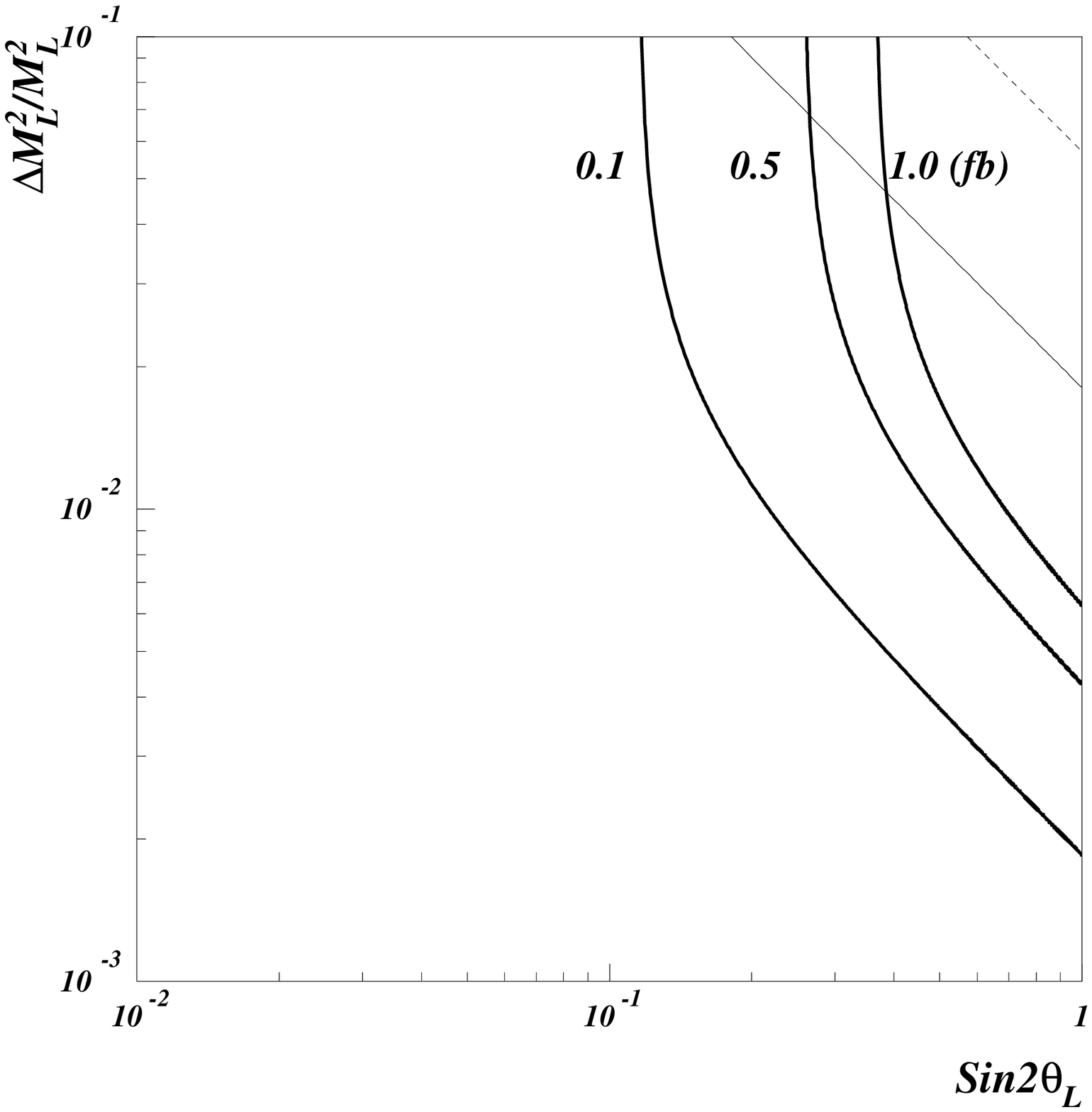}

\caption[]{Contours of constant cross sections for
$e^- \gamma \to  \tilde{\mu}_L \tilde{\chi}_2^0 \to 
\mu^- \Eslash\  q\bar{q}\ q \bar{q}$ 
(thick solid) with $ \sqrt{s_{ee}}=500 $ GeV and the SUSY 
parameters given in the text. 
Constant contours of $ B(\mu \to e \gamma ) =4.9 \times
10^{-11} $ (dotted) and $4.9 \times 10^{-12}$ (solid)
are also plotted.}
\end{figure}

\begin{figure}
\centering\leavevmode
\epsfxsize=5in\epsffile{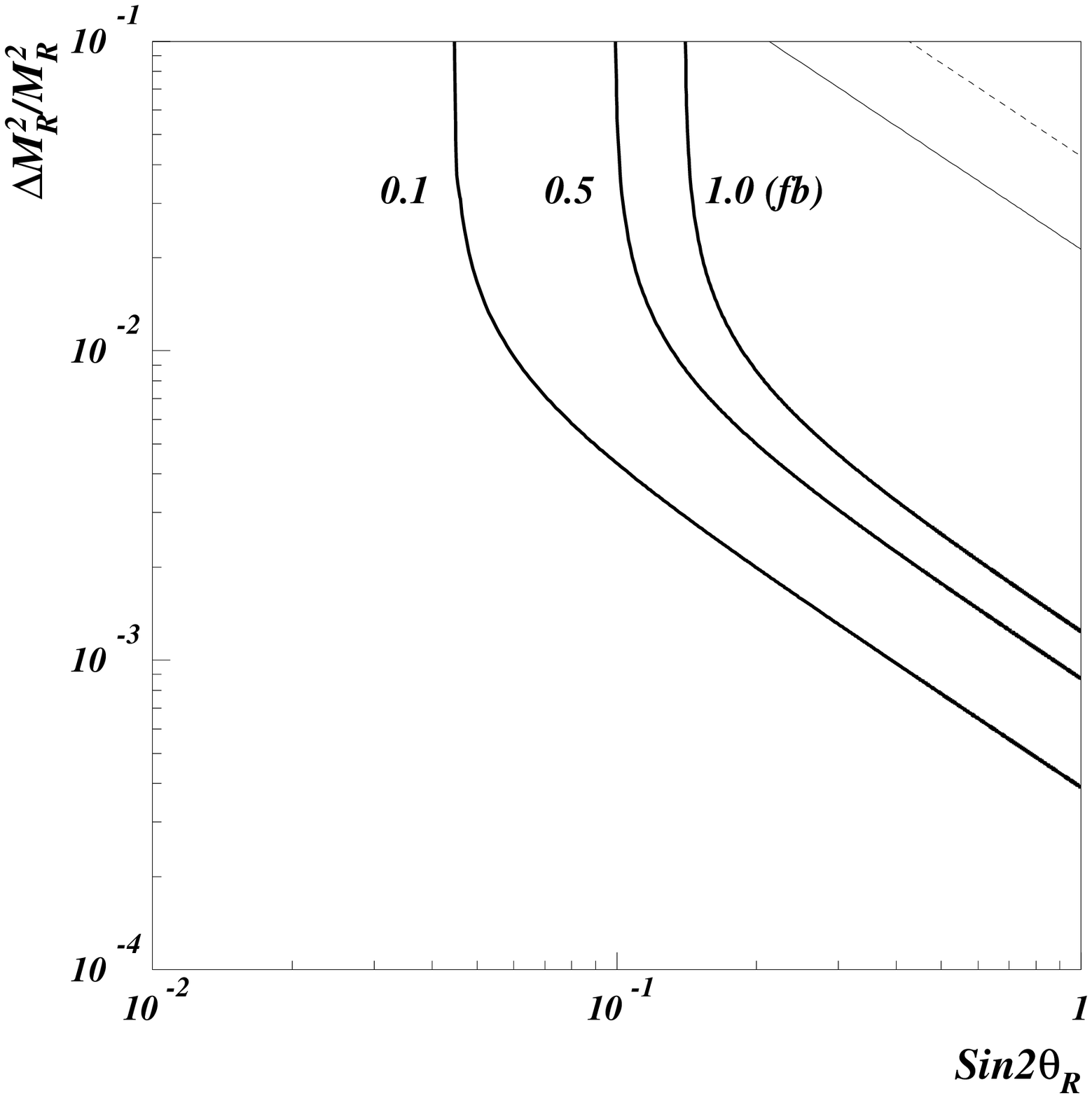}

\caption[]{Contours of constant cross sections for
$e_R^- \gamma_+ \to  \tilde{\mu}_R \tilde\chi_1^0 \to \mu^-\Eslash$
(thick solid) with $ \sqrt{s_{ee}}=500 $ GeV and the SUSY 
parameters given in the text. 
Constant contours of $ B(\mu \to e \gamma ) =4.9 \times
10^{-11} $ (dotted) and $4.9 \times 10^{-12}$ (solid)
are also plotted.}

\end{figure}

\end{document}